\documentstyle[12pt,epsfig]{article}
\def\sec{\,{\rm sec}}

\def\km{\,{\rm km}}

\def\Mpc{\,{\rm Mpc}}

\def\eV{{\,\rm eV}}

\def\cmm2{{\,\rm cm^{-2}}}
\def\cm2{{\,{\rm cm}^2}}
\def\cmm3{{\,{\rm cm}^{-3}}}
\def\gcmm3{{\,{\rm g\,cm^{-3}}}}

\def\mpl{{m_{\rm Pl}}} 

\def\la{\mathrel{\mathpalette\fun <}}

\def\fun#1#2{\lower3.6pt\vbox{\baselineskip0pt\lineskip.9pt
  \ialign{$\mathsurround=0pt#1\hfil##\hfil$\crcr#2\crcr\sim\crcr}}}

\begin{document}
\pagestyle{empty}
\begin{center}
\bigskip

\rightline{FERMILAB--Pub--98/247-A}
\rightline{astro-ph/9808133}
\rightline{to appear in {\it Physical Review D Rapids}}

\vspace{.2in}
{\Large \bf Prospects for probing the dark energy
via supernova distance measurements}
\bigskip

\vspace{.2in}
Dragan Huterer$^1$ and Michael S. Turner$^{1,2,3}$\\

\vspace{.2in}
{\it $^1$Department of Physics\\
Enrico Fermi Institute,
The University of Chicago, Chicago, IL~~60637-1433}\\

\vspace{0.1in}
{\it $^2$Department of Astronomy \& Astrophysics \\
Enrico Fermi Institute, The University of Chicago, Chicago, IL~~60637-1433}\\

\vspace{0.1in}
{\it $^2$NASA/Fermilab Astrophysics Center\\
Fermi National Accelerator Laboratory, Batavia, IL~~60510-0500}\\

\end{center}

\vspace{.3in}
\centerline{\bf ABSTRACT}
\bigskip

Distance measurements to Type Ia supernovae (SNe Ia)
indicate that the Universe is accelerating and that two-thirds
of the critical energy density exists in a dark-energy component with
negative pressure.  Distance measurements to SNe Ia can be
used to distinguish between different possibilities for
the dark energy, and if it is an evolving scalar
field, to reconstruct the scalar-field potential.
We derive the reconstruction equations and address the
feasibility of this approach by Monte-Carlo simulation.

\bigskip

\newpage
\pagestyle{plain}
\setcounter{page}{1}
\newpage

\section{Introduction}

There is now prima facie evidence that the Universe is
flat and that the critical energy density is 1/3 matter
and 2/3 something else with large, negative pressure.  The simplest
possibility for the latter component is vacuum energy
(cosmological constant) \cite{cc}; other possibilities include
a frustrated network of topological defects \cite{defects} and an
evolving scalar field \cite{sf1,sf2}, called
quintessence by the authors of Ref.~\cite{quint}.
All have effective bulk pressure that is very negative, $p\la -\rho /3$;
for the cosmological constant $p=-\rho$ and for a frustrated
defect network $p=-{N\over 3}\rho$ where $N$ is the dimension
of the defect.
In this paper we discuss the use of SNe Ia to distinguish between these
possibilities and to probe the scalar-field potential
associated with the quintessence field.

Backing up for a moment, the evidence for flatness comes from
measurements of the multipole power spectrum of the cosmic
background radiation (CBR) which show a peak around $l\simeq 200$
as expected for a flat Universe \cite{cbr_flat}.  A variety of
dynamical measurements of the mean matter density indicate that
$\Omega_M = 0.4\pm 0.1$ \cite{Omega_M}.  Recent measurements of
the distances to more than 50 SNe Ia out to redshift $z\sim 1$
indicate that the expansion is accelerating rather slowing down
\cite{SNeIa}.  If correct, this implies the existence of an
unknown component to the energy density with pressure $p_X \equiv
w_X\rho_X \la -\rho_X/3$ that contributes $\Omega_X \sim 0.6$
\cite{riess2}.  This fits neatly with the determinations that
$\Omega_M \sim 0.4$ and $\Omega_0$ $(= \Omega_X + \Omega_M) \sim
1$.  While this accounting is not yet definitive -- and could
possibly change dramatically -- it is worth thinking about how to
distinguish between the different possibilities suggested for the
unknown energy component \cite{tw}.

The key difference between quintessence and the other two
possibilities is that the effective equation of state,
$w_X=p_X/\rho_X$, can vary with time and can take on any value.
The combination of SNe Ia measurements and high-precision
measurements of the multipole power spectrum expected from the
MAP and Planck Surveyor satellites may be able to discriminate
between constant and varying $w_X$ \cite{ptw}.  If $w_X$ is found
to vary and/or is not equal to $-N/3$ ($N=1, 2, 3$), the next
question is how best to probe the ``quintessence sector.''  While
anisotropy of the CBR will be very powerful in determining many
important cosmological parameters, as we now explain, it has less
potential to probe the scalar-field potential than SNe Ia
measurements.  The fundamental reason is simple: CBR anisotropy
primarily probes the Universe at redshift $z\sim 1000$ when the
ratio of dark-energy density to matter density was tiny ($\ll
10^{-6}$); the SNe Ia probe the Universe at recent epochs when
the dark-energy density is beginning to dominate the matter
density.

Quintessence has three basic effects on CBR anisotropy.
The most significant is in determining the distance to
the last-scattering surface (Robertson--Walker coordinate
distance to redshift $z\simeq 1100$), which sets the geometric
relationship between angle subtended and length scale.
However, all models with the same distance to the last-scattering
surface will have essentially the same multipole power spectrum.
The second and third effects break this degeneracy, but
are less significant and/or powerful:  late-time ISW and
slight clumping of the scalar field (spatial inhomogeneity induced
by the lumpiness in the Universe) only affect
the lower-order multipoles, which can be less well determined
because of cosmic variance \cite{ptw}.

Supernovae on the other hand may be able to unravel
the essence of quintessence.  This is because accurate supernovae
distance measurements can map out $r(z)$ to redshift $z\sim 1$ or
perhaps higher, and this is when quintessence is becoming
dynamically important and where most of the ``scalar-field action''
is occurring.  [The quantity we focus on, coordinate distance to
redshift $z$, $r(z)$, is simply related to the quantity measured by
observers, luminosity distance, $d_L = (1+z)r(z)$.]  Shortly,
we will show the fact the scalar-field action
occurs at modest redshifts is a natural consequence of
quintessence.

In the next Section we will derive the reconstruction equations
for the scalar-field potential, and in the following Section 
we will address the practicality of this approach with
simulated data and Monte-Carlo realization of reconstruction.
We finish with a brief summary and concluding remarks.

\section{Reconstruction Equations}

We assume a flat Universe with two components to the
energy density:  nonrelativistic matter, which presently
contributes fraction $\Omega_M$ to the critical density,
and a single, homogeneous scalar field $\phi$ (for the problem
at hand, its slight clumping can be neglected).
The fundamental equations governing our cosmological model are:
\begin{eqnarray}
0 & = & \ddot\phi + 3H\dot\phi + V^\prime (\phi ) \\
r(z) & = & \int_{t_0}^{t(z)}\,du/a(u) = \int_0^z\,dx/H(x) \\
\left( {\dot a \over a}\right)^2 & \equiv & H(z)^2
        = {8\pi G\over 3}\rho  =
        {8\pi G\over 3} \left[ \rho_M + {1\over 2}{\dot\phi}^2
        + V(\phi ) \right] \nonumber \\
        & = & {1\over (dr/dz)^{2}} \\
\left( {\ddot a \over a} \right)
        & = & -{4\pi G\over 3}\left( \rho + 3p \right)
        = -{8\pi G\over 3}\left[ {1\over 2} \rho_M + {\dot\phi}^2
        - V(\phi ) \right] \nonumber \\
        & = & {1\over (dr/dz)^2}
        + (1+z){d^2r/dz^2 \over (dr/dz)^3}
\end{eqnarray}
where $r(z)$ is the Robertson--Walker coordinate distance to
an object at redshift $z$, the matter density $\rho_M =
\Omega_M \rho_{\rm crit} = 3\Omega_MH_0^2(1+z)^3/8\pi G$,
prime denotes derivative with respect to $\phi$, and the energy density
and pressure of the evolving scalar field are:
\begin{eqnarray}
\rho_\phi & = & {1\over 2}{\dot\phi}^2 + V(\phi ) \\
p_\phi    & = & {1\over 2}{\dot\phi}^2 - V(\phi )
\end{eqnarray}
Note too:  $dz/dt = -(1+z)H(z) = -(1+z)/(dr/dz)$ and $\Omega_\phi
\equiv \rho_\phi /\rho_{\rm crit}=1-\Omega_M$.  Since the relative fractions
of critical density in matter and quintessence evolve with time,
it is important to remember that $\Omega_M$ and $\Omega_\phi$ refer
to the present epoch.

As an aside, and before deriving the reconstruction equations, we
will show why quintessence models are likely to predict
interesting scalar-field dynamics recently.  As an indicator of
``interesting'' scalar-field dynamics, consider the time
derivative of the effective equation of state, $w_\phi
\equiv p_\phi /\rho_\phi$,
\begin{equation}
{d\ln w_\phi \over d\ln a} = 2{d\ln PE /d\ln a \over 1 - KE/PE}
        -12{KE\cdot PE\over KE^2-PE^2}
\label{eq:naturally}
\end{equation}
where $KE = {1\over 2}{\dot\phi}^2$, $PE = V(\phi )$, and the equality
follows from using the equation of motion for $\phi$.  If quintessence
is to be distinguishable from a cosmological constant, then $w_\phi$
must differ from $-1$, which implies that the kinetic and
potential terms are comparable.  Further, barring accidental
(or pre-arranged cancellations), Eq. (\ref{eq:naturally}) then
implies that $d\ln w/ d\ln a$ is presently of order unity.

On to reconstruction.  Since $r(z)$ is determined by $H(z)$ 
and $H(z)$ is a function
of the scalar field, one should be able to write down equations
for $V(\phi )$ and $\dot\phi$ in terms of $r(z)$.
The following is a parametric solution
for $V(\phi )$ and $d\phi /dz$, in terms of $r(z)$, $dr/dz$ and $d^2r/dz^2$:
\begin{eqnarray}
V[\phi (z)] & = & {1\over 8\pi G}\left[ {3\over (dr/dz)^2}
        +(1+z) {d^2r/dz^2\over (dr/dz)^3}\right]  - {3\Omega_MH_0^2
        (1+z)^3 \over 16\pi G} \\
{d\phi \over dz} & = & \mp {dr/dz \over 1+z}\, \left[ -{1\over 4\pi G}
        {(1+z)d^2r/dz^2 \over (dr/dz)^3} - {3\Omega_M H_0^2(1+z)^3\over
        8\pi G} \right]^{1/2}
\end{eqnarray}
where the upper (lower) sign applies if $\dot\phi >0$ ($<0$).
The sign in fact is arbitrary, as it can be changed by the
field redefinition, $\phi \leftrightarrow -\phi$.  The actual
value of $\phi$ cannot be determined by reconstruction:  $\phi$
can be shifted by an arbitrary constant with no cosmological effect
(the form of the potential of course changes).

In integrating the reconstruction equations it is useful
to define dimensionless quantities,
\begin{eqnarray}
x & \equiv & H_0t  \nonumber \\
\tilde r & \equiv & H_0 r \nonumber \\
\tilde\phi & \equiv & \phi /\mpl \nonumber \\
\tilde H & \equiv & H/H_0 = \sqrt{\Omega_M(1+z)^3 + \omega +{4\pi \over 3}
        \left({d\tilde\phi /dx }\right)^2 } \nonumber \\
\omega (\tilde\phi ) & \equiv & V({\tilde\phi}\mpl )/(3H_0^2/8\pi G)
\end{eqnarray}

The differential equations governing $\tilde r$, $\tilde\phi$ and $a$
become
\begin{eqnarray}
{da\over dx} & = & a \tilde H\\
{d\tilde r \over dz} & = & 1/\tilde H \\
0 & = & {d^2 \tilde \phi \over dx^2} + 3\tilde H {d\tilde\phi \over dx}
        +{3\over 8\pi} {d\omega \over d\tilde\phi}
\end{eqnarray}
and the construction equations are
\begin{eqnarray}
\omega (\tilde\phi (z)) & = & \left[ {1\over (d\tilde r/dz)^2}
        +{(1+z)\over 3}\,{d^2\tilde r/dz^2\over (d\tilde r/dz)^3}\right]
        -{1\over 2}\Omega_M(1+z)^3 \\
{d\tilde\phi \over dz} & = & \mp {d\tilde r/dz\over (1+z)}
        \left[ -{1\over 4\pi}{(1+z)d^2\tilde r/dz^2\over (d\tilde r/dz)^3}
        -{3\over 8\pi}\,\Omega_M(1+z)^3\right]^{1/2}
\end{eqnarray}
The boundary conditions, expressed at the present epoch, are:
$\tilde H_0=1$, $0 < \omega (\tilde\phi_0) < 1-\Omega_M$
and $d\tilde \phi /dx = \sqrt{{3\over 4\pi} [1-\Omega_M -\omega ]}$.

Finally, without recourse to a scalar-field model for the
unknown, negative pressure component, one can derive
a reconstruction equation for the bulk equation of
state, $w_X = p_X/\rho_X$, as a function of redshift.
The equation of motion for the X-component,
\begin{equation}
d\ln \rho_X = -3(1+w_X)d\ln (1+z) \qquad\Rightarrow\qquad
        \rho_X (z) = {3H_0^2\over 8\pi G}\,e^{3\int_0^z\,(1+w_X)d\ln (1+z)}\,,
\end{equation}
is used in place of Eq.~(1), and the reconstruction equation
for $w_X(z)$ is derived as before,
\begin{equation}
1+w_X(z) = {1+z\over 3}\, {3H_0^2\Omega_M(1+z)^2 + 2(d^2r/dz^2)/(dr/dz)^3\over
        H_0^2\Omega_M(1+z)^3-(dr/dz)^{-2}}
\label{eq:w-recon}
\end{equation}
Using Eq.~(\ref{eq:w-recon}), SNe Ia measurements alone can be
used to determine $w_X$ and address its time variation.

\section{Simulating Reconstruction}

\begin{figure}[ht]
\centerline{\epsfig{figure=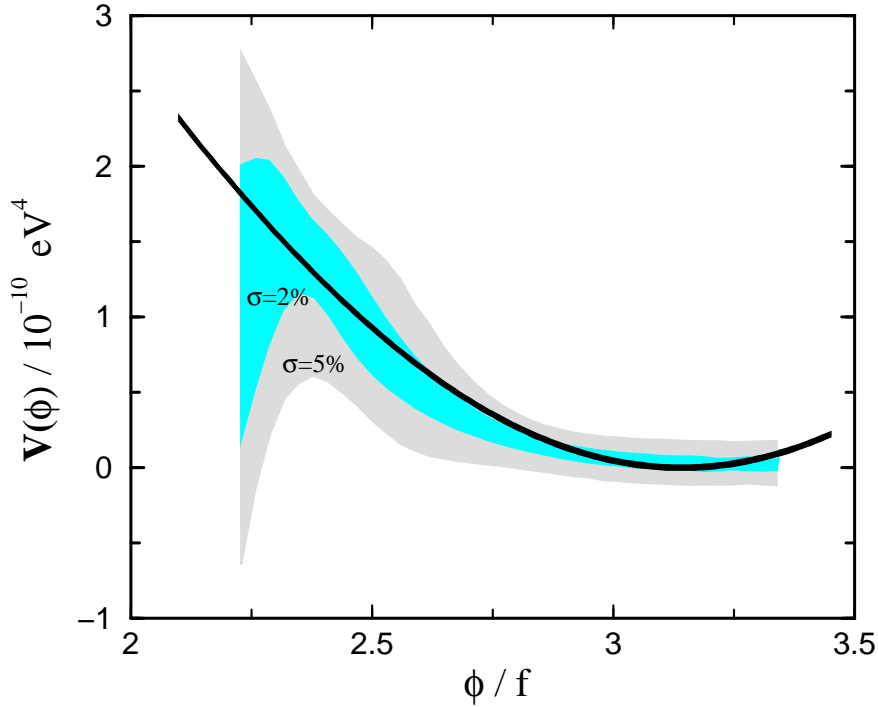,width=4.5in}}
\caption{The 95\% confidence interval for the reconstructed
potential assuming luminosity distance errors of 5\% and 2\%
(shaded areas) and the original potential (heavy line).
For this reconstruction, $\Omega_M = 0.3$,
$V(\phi ) = V_0[1+\cos (\phi /f) ]$,
$V_0 = (4.65\times 10^{-3}\eV )^4$, $f/\mpl = 0.154$,
$N=40$, and $z_{\rm max} = 1.0$.  The simulated data
were fit by a fourth-order polynomial in $z$.
}
\label{fig:recon1}
\end{figure}

Here we investigate the feasibility of our approach and estimate the
inherent errors by means of simulated data and Monte-Carlo
realization.  Our procedure is straightforward:
\begin{itemize}

\item Pick a potential $V(\phi)$, matter density $\Omega_M$, 
and values for $\phi_0$ and $\dot\phi_0$ (consistent with
$\Omega_\phi = 1 - \Omega_M$)

\item Compute the evolution of $\phi$, $a(t)$ and $r(z)$ for this
quintessence model by evolving $\phi (t)$ and $a(t)$ back in time

\item Realize the model by simulating SNe Ia measurements:
for $z_i = (i-1)\Delta z$ and $i=1$ to $N$, $r_i = r(z_i) +
\delta r_i$ where $\delta r_i$ is drawn from a Gaussian distribution
with zero mean and variance $\sigma\, r_i(z)$ ($\sigma$
is the relative error in the luminosity distance)

\item Fit the simulated data with a (low-order) polynomial and 
numerically compute $V(\phi )$ from the reconstruction equations

\item Repeat one thousand times to estimate the error in
reconstructing $V(\phi )$

\item This procedure actually reconstructs $\omega (\phi)$; to
get $V(\phi)$ one simply takes a value for $H_0$; for our results 
we took $H_0=70 \km \,\sec^{-1} \Mpc^{-1}$
\end{itemize}

\begin{figure}[ht]
\centerline{\epsfig{figure=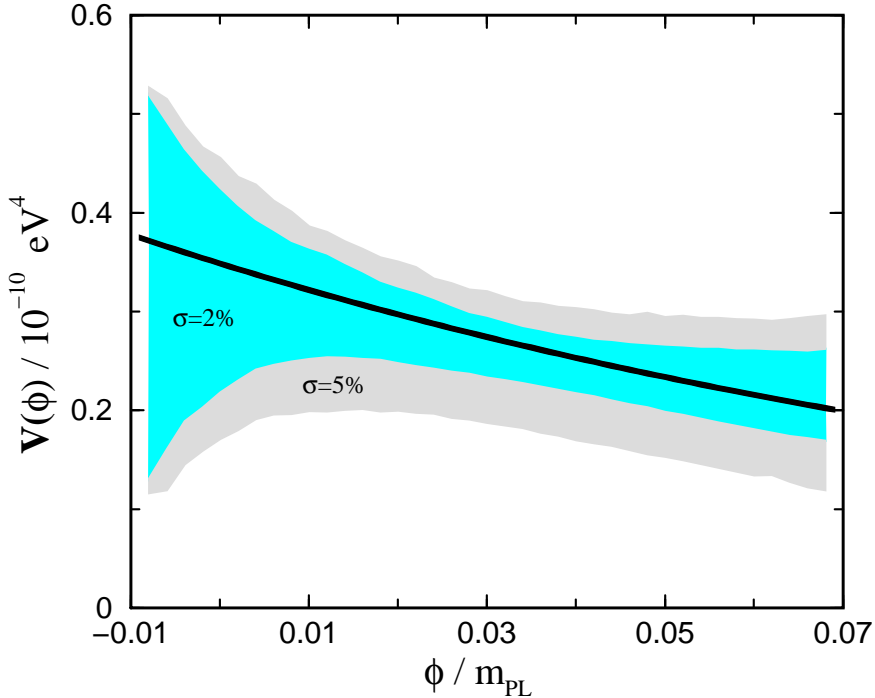,width=4.5in}}
\caption{The 95\% confidence interval for the reconstructed potential
assuming luminosity distance errors of 5\% and 2\% (shaded areas) and the
original potential (heavy line).  For this reconstruction, $\Omega_M = 0.4$,
$V(\phi ) = V_0\exp (-\beta \phi /\mpl )$,
$V_0 = (2.43\times 10^{-3}\eV )^4$, $\beta = 8$,
$N=40$, and $z_{\rm max} = 1.5$.  The simulated data
were fit by a fourth-order polynomial in $z$.
}
\label{fig:recon2}
\end{figure}

Before presenting some results, we should elaborate on a few
technical details. In fitting a polynomial to $r_i$ we have tried
third-, fourth-, fifth-, and sixth-order polynomials; all give
similar results.  Because we are taking derivatives of $r(z)$,
the use of higher-order polynomials only introduces numerical
``noise'' and is not useful.  We have varied the number of
redshift bins $N$ from 20 to 100, and $z_{\rm max}\equiv
(N-1)\Delta z$ from 1 to 1.5; the errors in the reconstruction
scale roughly as expected, $1/\sqrt{N}$.  We also tried using a
Gaussian distribution in redshift (with more data points at small
$z$ than at high $z$); the results change little relative to the
case of linear distribution.

We have reconstructed several potentials; here, we present
results for the exponential and cosine potentials
considered previously \cite{sf2,quint}.  Shown in Figs.\ 1 and 2
are the original potential and the 95\% confidence intervals for
the reconstructed potential for data with 5\% and 2\% luminosity
distance errors.  The confidence intervals are obtained by
requiring that 950 of the 1000 Monte Carlo realizations give a
value of the potential in that interval. The error in the
reconstructed potential is mostly attributable to the $d^2r/dz^2$
terms in the reconstruction equations, reflecting the fact that
it is extremely difficult to infer the second derivative of the
noisy data.  In particular, an uncertainty of 10\%
in $\Omega_M$ does not change the reconstruction
confidence regions appreciably.

\begin{figure}[ht]
\centerline{\epsfig{figure=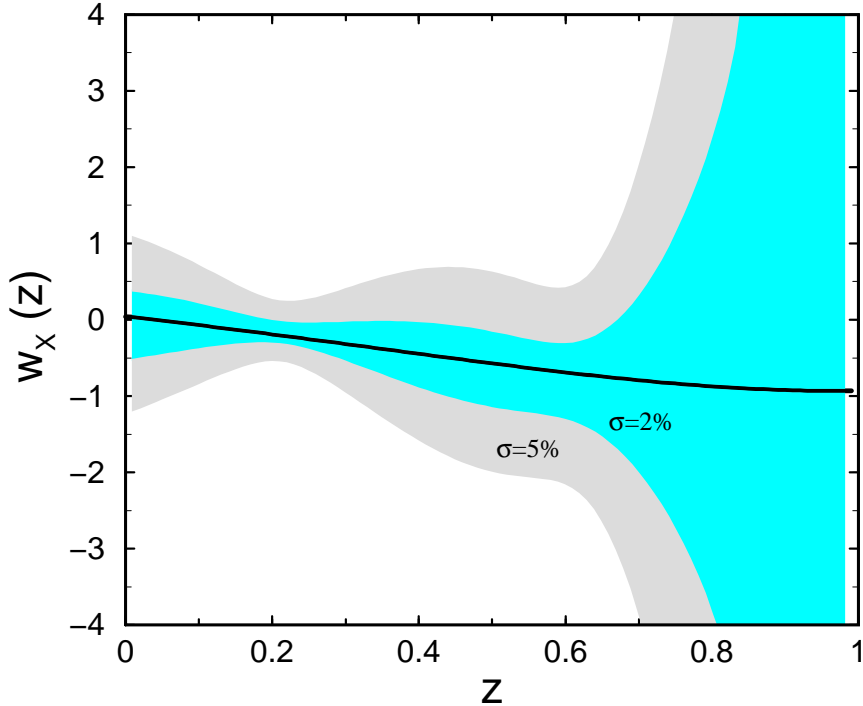,width=4.5in}}
\caption{The 95\% confidence interval for the reconstructed
equation of state of the unknown component assuming luminosity
distance errors of 5\% and 2\% (shaded areas) and the original
equation of state (heavy line).  For this reconstruction,
$\Omega_M =0.4$, $V(\phi ) = V_0\exp (-\beta \phi /\mpl )$, $V_0
= (2.43\times 10^{-3}\eV )^4$, $\beta = 15$, $N=40$, and $z_{\rm
max} = 1.0$.  The simulated data were fit by a  third-order
polynomial in $z$. } 
\label{fig:recon3}  
\end{figure} 

Finally, Fig.\ \ref{fig:recon3} shows the reconstruction of the
equation of state for an exponential potential.  Even though
the uncertainties in the reconstruction are large, we are still
able to distinguish this quintessence model from a constant equation
of state.  In particular, using the $\chi^2$ statistic for $r(z)$
and the simulated data with 2\% errors, {\it all} constant
equations of state can be ruled out with 99.9\% confidence, and
the most plausible constant equations of state, $w_X = -N/3$ for $N=0,1,2,3$,
can be ruled out with more than 99.9\% confidence.
(Of course, the ability to discriminate
between constant $w_X$ and varying $w_X$ depends upon how
$w_X$ varies.)  Fig.\ \ref{fig:recon3} also nicely illustrates
a general feature of reconstruction:  Because the fractional
contribution of dark energy rapidly decreases with increasing
redshift, $\rho_X/\rho_M \propto (1+z)^{3w_X}$, reconstruction
beyond redshift of around $z\sim 0.8$ becomes extremely difficult.
Thus, for the purposes of reconstruction, data at redshifts
greater than unity are of very limited value.

\section{Discussion}

If correct, the discovery that the expansion of the Universe is
speeding up rather than slowing down is one of the most important
discoveries of this century.  It implies that the primary component
of the cosmos today is dark energy with large negative pressure and
of unknown composition.  The implications for fundamental physics
are equally profound as all the possibilities for the dark-energy
component are deeply rooted in fundamental physics:
vacuum energy, a frustrated network of topological defects and
an evolving scalar field.

The measurement of distances to supernovae out to redshift of order unity,
which led to this discovery, may also be of great utility
in determining the character of the dark energy.
In this paper we have derived the equations that relate the equation
of state of the dark-energy component and, in the case of quintessence, the
scalar-field potential to measurements of the luminosity distance.
By use of Monte-Carlo simulation we have shown how this might be done
in practice and addressed the feasibility of this technique.

While ours is a preliminary investigation, the results indicate that
using SN Ia measurements to probe the dark-energy component is
promising.  However, important questions and issues remain
before one can be confident that this technique can be used in
practice.  Some involve the technical details of how our method
might be implemented:  What is the optimal distribution
of supernova redshifts for probing $w_X$ and $V(\phi )$?
Would a likelihood analysis for a parameterized fit to the
potential be more powerful than our reconstruction approach?
We have already begun to address these questions with some
success.  For example, we have found that Pade approximants
are a much better way to represent the observational data
than polynomials (splines may do even better); for many potentials
a linear distribution in redshift minimizes the area of the
95\% confidence region for the reconstructed potential.

Foremost among the open issues is the reliability of SNe Ia
as distance indicators.  Currently, the distances errors are estimated to be
of order 10\% to 20\% (per supernova) \cite{schmidt}.
They arise from a variety of sources:  reddening due to the host
galaxy or intergalactic dust; intrinsic dispersion in the brightness -- decline
relationship (Phillips relation \cite{phillips}) used to
calibrate the SNe Ia; possible systematic (with redshift) evolutionary effects;
and possible dependence upon the differing chemical composition
of the supernova progenitors.  There is much activity,
both theoretical and observational, directed at better understanding
type Ia supernovae and so we can hope for improvement in
reducing and/or better understanding systematic errors.  Further,
if the intrinsic scatter (both statistical and systematic)
in the brightnesses of SNe Ia is random
(not correlated with redshift), then distance error can be beaten down by
more measurements.  For example, allowing 10\% distance error
per supernova, 3\% errors in $r(z)$ could be obtained if 10 supernovae
are measured at each redshift, increasing the total needed for
our reconstruction method to 500 or so.  Finally, the two
groups have proven that discovering 1000s of SNe1a over the next
decade is a very realistic goal, and with a better
understanding of SNe Ia, one could cull a large sample to form
a smaller, higher quality sample with smaller systematic error and
statistical scatter.

To summarize, distance measurements to type Ia supernovae with
redshifts less than order unity
have the potential to shed light upon the nature of
the dark energy.  Based upon existing SNe Ia measurements and
their uncertainties,
it appears that obtaining data of the quantity and quality
required to probe the dark-energy component using the method
we have described, while not guaranteed, is also not unrealistic.

\paragraph{Acknowledgments.}  This work was supported by the DoE (at
Chicago and Fermilab) and by the NASA (at Fermilab through grant NAG
5-7092).  MST thanks the Aspen Center for Physics for providing a
quiet, but stimulating place to finish this work.

\end{document}